\documentclass[sn-basic,Numbered]{sn-jnl}% Basic Springer Nature Reference Style/Chemistry Reference Style
\usepackage{graphicx}%
\usepackage{multirow}%
\usepackage{amsmath,amssymb,amsfonts}%
\usepackage{amsthm}%
\usepackage{mathrsfs}%
\usepackage[title]{appendix}%
\usepackage{xcolor}%
\usepackage{textcomp}%
\usepackage{manyfoot}%
\usepackage{booktabs}%
\usepackage{algorithm}%
\usepackage{algorithmicx}%
\usepackage{algpseudocode}%
\usepackage{listings}%
\theoremstyle{thmstyleone}%
\theoremstyle{thmstyletwo}%

\theoremstyle{thmstylethree}%

\usepackage{verbatim}
\newcommand{\detailtexcount}[1]{%
  \immediate\write18{texcount -merge -sum -q sn-article.tex sn-bibliography.bib > sn-article.wcdetail }%
  \verbatiminput{sn-article.wcdetail}%
}

\newcommand{\quickwordcount}[1]{%
  \immediate\write18{texcount -1 -sum -merge -q sn-article.tex sn-bibliography.bib > sn-article-words.sum }%
  \input{sn-article-words.sum} words%
}

\newcommand{\quickcharcount}[1]{%
  \immediate\write18{texcount -1 -sum -merge -char -q sn-article.tex sn-bibliography.bib > sn-article-chars.sum }%
  \input{sn-article-chars.sum} characters (not including spaces)%
}

\begin{document}

\title[Bayesian Active Learning in MSK Segmentation]{Hybrid Representation-Enhanced Sampling for Bayesian Active Learning in Musculoskeletal Segmentation of Lower Extremities}


\author*[1]{\fnm{Ganping} \sur{Li}}\email{li.ganping.lc2@is.naist.jp}

\author[1]{\fnm{Yoshito} \sur{Otake}}\email{otake@is.naist.jp}

\author[1]{\fnm{Mazen} \sur{Soufi}} \email{msoufi@is.naist.jp}

\author[2]{\fnm{Masashi} \sur{Taniguchi}}\email{taniguchi.masashi.7a@kyoto-u.ac.jp}

\author[2]{\fnm{Masahide} \sur{Yagi}}\email{yagi.masahide.5s@kyoto-u.ac.jp}

\author[2]{\fnm{Noriaki} \sur{Ichihashi}}\email{ichihashi.noriaki.5z@kyoto-u.ac.jp}

\author[3]{\fnm{Keisuke} \sur{Uemura}}\email{surmountjp@gmail.com}

\author[4]{\fnm{Masaki} \sur{Takao}}\email{takao.masaki.ti@ehime-u.ac.jp}

\author[3]{\fnm{Nobuhiko} \sur{Sugano}}\email{n-sugano@umin.net}

\author[1]{\fnm{Yoshinobu} \sur{Sato}}\email{yoshi@is.naist.jp}

\affil*[1]{\orgdiv{Division of Information Science, Graduate School of Science and Technology}, \orgname{Nara Institute of Science and Technology}, \orgaddress{\street{8916-5 Takayama}, \city{Ikoma}, \postcode{630-0192}, \state{Nara}, \country{Japan}}}

\affil[2]{\orgdiv{Human Health Sciences, Graduate School of Medicine}, \orgname{Kyoto University}, \orgaddress{\street{53-Kawahara-cho, Shogoin}, \city{Sakyo-ku}, \postcode{606-8507}, \state{Kyoto}, \country{Japan}}}

\affil[3]{\orgdiv{Department of Orthopedic Surgery, Osaka University Graduate School of Medicine}, \orgname{Osaka University}, \orgaddress{\street{2-2 Yamadaoka}, \city{Suita}, \postcode{565-0871}, \state{Osaka}, \country{Japan}}}

\affil[4]{\orgdiv{Department of Bone and Joint Surgery, School of Medicine}, \orgname{Ehime University}, \orgaddress{\street{454 Shitsugawa}, \city{Toon}, \postcode{791-0295}, \state{Ehime}, \country{Japan}}}

%------------------------- Text Count -------------------------------
% Don't count these!
% TC:ignore
% \quickwordcount{sn-article}
% \quickcharcount{sn-article}
% \detailtexcount{sn-article}
% TC:endignore

%--------------------------------------------------------------------


\abstract{Purpose: Manual annotations for training deep learning (DL) models in auto-segmentation are time-intensive. This study introduces a hybrid representation-enhanced sampling strategy that integrates both density and diversity criteria within an uncertainty-based Bayesian active learning (BAL) framework to reduce annotation efforts by selecting the most informative training samples.

Methods: The experiments are performed on two lower extremity (LE) datasets of MRI and CT images, focusing on the segmentation of the femur, pelvis, sacrum, quadriceps femoris, hamstrings, adductors, sartorius, and iliopsoas, utilizing a U-net-based BAL framework. Our method selects uncertain samples with high density and diversity for manual revision, optimizing for maximal similarity to unlabeled instances and minimal similarity to existing training data. We assess the accuracy and efficiency using Dice and a proposed metric called reduced annotation cost (RAC), respectively. We further evaluate the impact of various acquisition rules on BAL performance and design an ablation study for effectiveness estimation.

Results: In MRI and CT datasets, our method was superior or comparable to existing ones, achieving a 0.8\% Dice and 1.0\% RAC increase in CT (statistically significant), and a 0.8\% Dice and 1.1\% RAC increase in MRI (not statistically significant) in volume-wise acquisition. Our ablation study indicates that combining density and diversity criteria enhances the efficiency of BAL in musculoskeletal segmentation compared to using either criterion alone.

Conclusion: Our sampling method is proven efficient in reducing annotation costs in image segmentation tasks. The combination of the proposed method and our BAL framework provides a semi-automatic way for efficient annotation of medical image datasets.}




\maketitle

\section{Introduction}\label{sec1}

Medical image segmentation is crucial in extracting quantitative imaging markers for better observations of anatomical or pathological structure changes. Its automation is essential not only for computer-aided musculoskeletal disease \cite{loureiro2013muscle, uemura2016volume} but also in investigating atrophy and fatty degeneration of individual muscles due to aging or disease (e.g. osteoarthritis) progression using a large number of data \cite{ogawa2020validation, yagi2022age}. However, obtaining manual annotations for training deep learning (DL) models is often time-consuming, resulting in insufficient model performance \cite{sourati2018active}. Active learning (AL) is a widely-adopted approach to address the above-mentioned issue \cite{settles2008analysis}. The method is regarded as a training schema that reduces annotation effort by sequentially annotating the most informative instances. It involves iterative steps where an AL framework selects a batch of samples from an unlabeled pool to be manually annotated by annotators and subsequently added to the training pool. Afterward, a new model is trained on the updated training pool, superseding the previous model in the framework. Though many recent studies have proposed competitive strategies to address the issue, the best sampling policy is still a matter of debate \cite{budd2021survey}. 

Prior AL approaches focus on selecting samples with high model uncertainty \cite{gal2016dropout,gal2017deep,lakshminarayanan2017simple, GAILLOCHET2023102958}, which is called uncertainty-based sampling. Among the uncertain samples, two criteria have been used for further selection \cite{yang2017suggestive,hiasa2019automated, ozdemir2021active, smailagic2018medal, nath2020diminishing, li2023hal}. One criterion \cite{yang2017suggestive,hiasa2019automated} seeks to select representative samples of high-density dominant classes in data distribution by maximizing their similarity to the unlabeled data. Alternatively, the others \cite{smailagic2018medal,nath2020diminishing, li2023hal} select samples minimizing the similarity between the chosen samples and the existing labeled data, ensuring diversity and less redundancy. To our knowledge, while the integration of density and diversity criteria with uncertainty-based methods has been applied to classification tasks in various fields \cite{liu2022survey}, their application in medical image segmentation has not been previously explored.

In this study, we introduce an AL scheme incorporating the two criteria above to ensure the chosen samples' representativeness and the labeled data's diversity. In order to further identify the informative instances, we implement a Bayesian active learning (BAL) framework based on Bayesian U-net \cite{hiasa2019automated} for uncertainty estimation and sampling. Since CT and MR data typically consist of volumetric (3D) images, volume-wise sample acquisition is preferable. However, the performance of AL approaches on volume segmentation tasks is relatively undercharacterized as prior research has mainly addressed 2D segmentation tasks, except for \cite{nath2020diminishing, ozdemir2021active}. Therefore, we will assess our proposed approach on both 2D and 3D segmentation tasks. In brief, our contributions can be summarized as follows:

\begin{quote}
    \begin{itemize}
        \item We proposed a hybrid representation-enhanced sampling strategy that integrates similarity measures to detect high-density samples while ensuring diversity and less redundancy. The method is adopted to a BAL framework to prioritize both uncertainty and representativeness of the queried samples for medical image segmentation.
        \item We validate our method on two lower extremity (LE) image datasets of MRI and CT, and further estimate the impact of the volume-wise acquisition in addition to the slice-wise on BAL performance and annotation efficiency, addressing the insufficiency in previous works.
    \end{itemize}
\end{quote}

\section{Related work}\label{sec2}

\subsection{Uncertainty-based sampling}\label{subsec1}
Uncertainty-based sampling assesses a sample's informativeness by measuring the uncertainty of a trained DL model's prediction, where a higher uncertainty indicates greater informativeness. Gal et al. \cite{gal2016dropout,gal2017deep} introduced a widely used implementation to approximate Bayesian inference using Monte Carlo (MC) dropouts. The method efficiently estimates model uncertainty by measuring each test sample's degree of difference at the inference step, originating from the deficiency of training data \cite{hiasa2019automated}. Nevertheless, given that a model in an early stage of AL tends to be uncertain for similar types of instances, relying solely on uncertainty approaches may skew the model to focus on a particular area of the data distribution within the target domain \cite{yang2017suggestive, budd2021survey, li2023hal}. Addressing this issue, Gailllochet et al. \cite{GAILLOCHET2023102958} proposed an uncertainty-based stochastic batch querying method. This approach aims to enhance the diversity within each batch of uncertain samples, aligning with the strategies discussed in the following section.

\subsection{Representativeness-based sampling}\label{subsec2}
Representativeness-based sampling is widely employed with uncertainty approaches \cite{budd2021survey} in medical analysis, mainly grouped by density-based and diversity-based approaches. As a typical density method, Yang et al. \cite{yang2017suggestive} utilized similarity measures to select dense samples that offer the most comprehensive representation of the unlabeled pool with a step-by-step optimization. In \cite{ozdemir2021active}, a variational autoencoder (VAE)-based density sampling was proposed for the same purpose, although it requires an auxiliary model. These methods, however, might skew the training pool by selecting only the majority when handling an imbalanced dataset, especially in the early iterations. In order to tackle this challenge, diversity-based approaches have been proposed. Smailagic et al. \cite{smailagic2018medal} quantified the dissimilarity between feature maps of the chosen samples and the training pool, intending to maximize this dissimilarity. Then, Nath et al. \cite{nath2020diminishing} adopted mutual information as a regularizer to ensure diversity in training data with a similar aim. However, solely maximizing diversity may result in querying outliers \cite{amagata2023diversity}, which often signify rare or extreme cases unrepresentative of general data patterns. This risks the model overfitting to atypical instances, resulting in poor performance on typical data and possibly introducing bias or inaccuracies due to noise or errors in outliers. Thus, developing a hybrid sampling strategy that maximizes the density and diversity of the training data would be necessary.

\section{Materials and methods}\label{sec3}

\subsection{Dataset}\label{subsec3}

\begin{figure}[t]%
\centering
\includegraphics[width=0.9\textwidth]{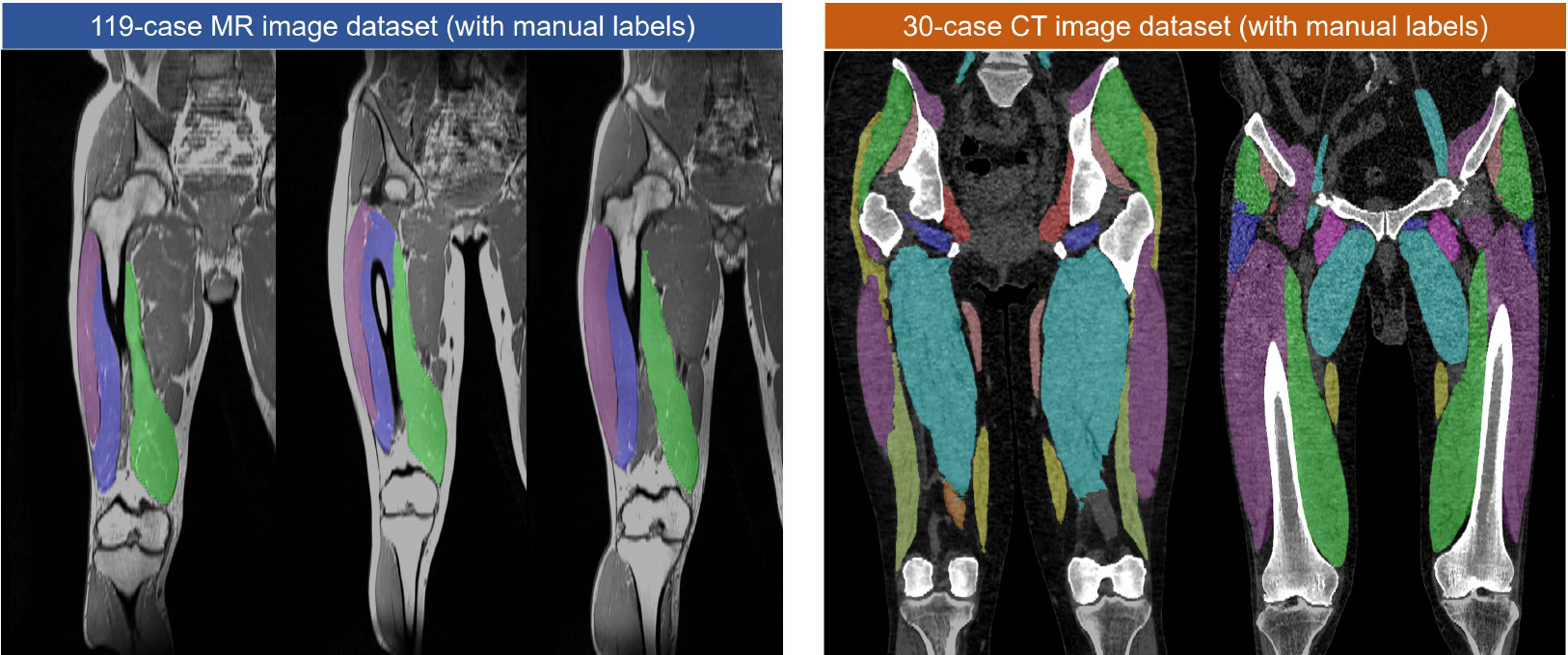}
\caption{Coronal views of T-1 weighted MR volumes (left) annotated with four quadriceps muscles, and CT volumes (right) annotated with 22 musculoskeletal structures.}\label{fig1}
\end{figure}

We gathered and annotated two lower extremity (LE) datasets (Fig. \ref{fig1}) used in our studies; 1) a T-1 weighted MRI dataset with four quadriceps muscles annotated \cite{fukumoto2022influence} and 2) a CT dataset with 22 musculoskeletal structures (19 muscles and 3 bones) annotated. The study received ethical approval from Osaka University (IRB approval no. 21115), Kyoto University (IRB approval no. R1746-2), and Nara Institute of Science and Technology (IRB approval no. 2020-M-7).

1) MRI dataset: The dataset consists of 119 MRI images (21,490 axial slices) from two age groups: 82 older people aged 60-89, recruited following a fitness examination in Kyoto, and 37 younger people aged 20-39 from Kyoto University. All participants were independent and ambulatory, with no contraindications for MRI. Labels for the four quadriceps muscles (rectus femoris, vastus lateralis, vastus intermedius, and vastus medialis) were initially created by three annotators using OsiriX \cite{rosset2004osirix}, and subsequently reviewed and revised by a medical expert. Each MR image contains 163 to 201 (182 on average) slices. The images were resized to $256\times256\times n$ with voxel spacing from $[0.5, 0.5, 4]$ mm to $[1, 1, 4]$ mm, and normalized from $[0, 1000]$ to $[0, 1]$. The dataset was then divided into 1/89/9/20 for the training/unlabeled pool/validation/testing for initialization of all AL experiments.

2) CT dataset: This dataset includes 30 preoperative CT images (17909 axial slices) from patients with unilateral hip joint osteoarthritis. Annotations for 22 musculoskeletal structures, including the femur, pelvis, sacrum, quadriceps femoris, hamstrings, adductors, sartorius, and iliopsoas, were initially generated using a pre-trained model \cite{hiasa2019automated}. Subsequently, these annotations were refined and verified by an intermediate-level orthopedic surgeon and a medical physicist in musculoskeletal imaging, utilizing 3D Slicer \cite{kikinis20133d} for the corrections. Each CT image contained 526 to 700 (599 on average) slices with a matrix size of $512\times512$. We resized the images to $256\times256\times n$ with voxel spacing of $[1.4, 1.4, 1]$ mm, normalized them from $[-150, 350]$ to $[0, 1]$, and divided the dataset into 1/24/1/4 for the training/unlabeled pool/validation/testing.

We focused on maximizing the size of the unlabeled pool to better estimate the algorithm’s performance across a diverse range of data. This was crucial given the limited number of images available. Consequently, our validation set was relatively small, a necessary trade-off to ensure a comprehensive representation in the unlabeled pool. Note that the manual labels of the unlabeled pool were used for evaluation only. During experiments, any slice or volume selected for manual revision uses its pre-existing label, which is treated as expert-revised and included in the training pool.

\subsection{Sampling techniques}\label{subsec4}
In this section, we present sampling techniques employed within our BAL framework (Fig. \ref{fig2} (a)). We start by introducing uncertainty sampling to select the most uncertain samples. Next, we incorporate a hybrid scoring approach designed to select high-density and diverse samples from the uncertain subset, thereby ensuring representativeness among the unlabeled data.

\textbf{Bayesian uncertainty sampling.}
Our uncertainty estimation step follows the method described in \cite{hiasa2019automated}, which investigates the model uncertainty in a scalable manner by the approximate Bayesian inference of predictive distributions (details shown in Online Resource 1.1). We implemented a dropout-based Bayesian U-net for multi-class segmentation and uncertainty estimation, where an average uncertainty for each class is defined by
\begin{equation}\label{eq1}
    m_{unc}(y=l) = \frac{1}{N}\sum\limits_{n=1}^{N}\text{var}[p(y=l\mid x, \Theta_{t})^{(n)}]
\end{equation}
where $N$ is the number of pixels of input $x$ and var$[p(y=l\mid x, \Theta_{t})^{(n)}]$ indicates the prediction variance under $T$ times Bayesian inference at pixel $n$. The equation above is cited and summarized from \cite{ozdemir2021active}. 

\textbf{Hybrid representation-enhanced sampling.}
We introduce a hybrid scoring approach to select high-density samples following the method described in \cite{yang2017suggestive,hiasa2019automated} with a constraint to maintain diversity \cite{nath2020diminishing}. 

Given an unlabeled pool $\mathcal{D}_u$, a training pool $\mathcal{D}_t$, and a subset of uncertain images $\mathcal{D}_{c}\subseteq\mathcal{D}_{u}$, the algorithm first measures the norm of cosine similarity, $norm(sim(I_{i}^{c}, I_{j}^{u}))$, between $\mathcal{D}_{c}$ and $\mathcal{D}_{u}$ for representative samples, where $I_{i}^{c}$ and $I_{j}^{u}$ are the $i^{th}$ and $j^{th}$ images from $\mathcal{D}_{c}$ and $\mathcal{D}_{u}$, respectively. Next, a regularization term of the mutual information's norm, $norm(mi(I_{i}^{c}, I_{k}^{t}))$, between $\mathcal{D}_{c}$ and $\mathcal{D}_{t}$ minimizes the redundancy in the training pool $\mathcal{D}_{t}$ while encouraging minority samples. Overall, we select samples maximizing
\begin{equation}\label{eq2}
    m_{repr} = \underbrace{norm(sim(\mathcal{D}_{c}, \mathcal{D}_{u}))}_\text{density module} - \lambda\cdot\underbrace{norm(mi(\mathcal{D}_{c}, \mathcal{D}_{t}))}_\text{diversity module}
\end{equation}
where hyper-parameter $\lambda$ determines the balance between the density and diversity of the chosen samples. The top-$k$ samples will then be annotated and added to the training pool $\mathcal{D}_{t}$. Details of the step-by-step optimization algorithm extended from \cite{hiasa2019automated, nath2020diminishing} are demonstrated in Online Resource 1.2.

\subsection{Bayesian active learning}\label{subsec6}
\begin{figure}[t]%
\centering
\includegraphics[width=0.95\textwidth]{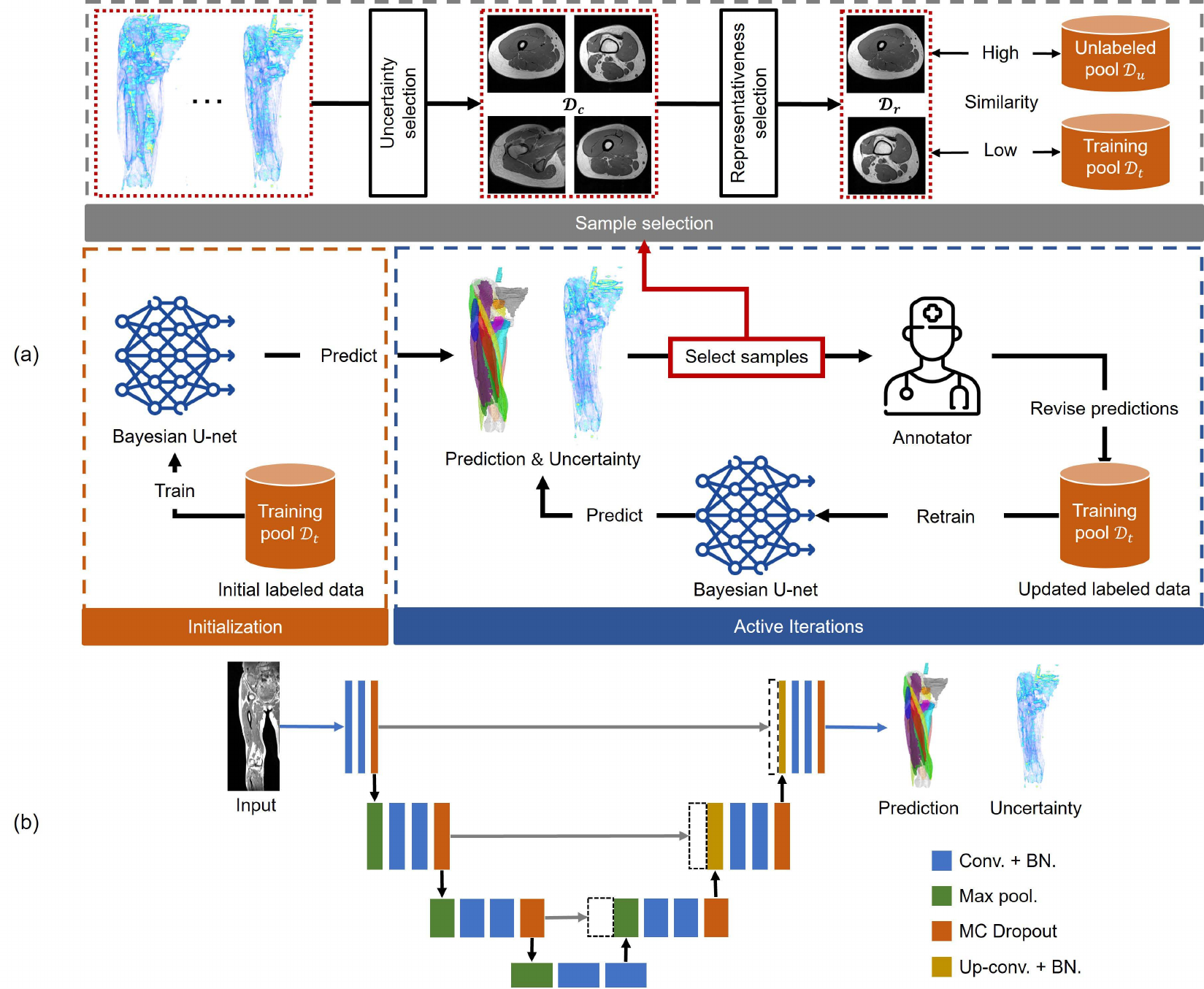}
\caption{Experiment layout. (a) Workflow of our BAL framework. Selected from the uncertain slices $\mathcal{D}_{c}$, the chosen samples $\mathcal{D}_{r}$ are intended to be representative of $\mathcal{D}_{u}$ while maintaining low MI with $\mathcal{D}_{t}$. (b) The architecture of our Bayesian U-net.}\label{fig2}
\end{figure}

We present a BAL framework for medical image segmentation to validate the proposed method. As shown in Fig. \ref{fig2} (a), we start with a Bayesian U-net trained on a limited number of labeled data, and then our schema iteratively selects uncertain and representative samples. These selected samples are treated as having been revised by annotators, utilizing their pre-existing labels, and are then incorporated into the training set.

\textbf{Segmentation model.}
Our segmentation tasks are conducted on a 4-layer Bayesian U-net \cite{hiasa2019automated} of 2.73 million trainable parameters whose architecture is depicted in Fig. \ref{fig2} (b). To tackle class imbalance, we employ a multi-class focal loss \cite{lin2017focal} with class weighter $\alpha=0.67$ and regularizer $\gamma=2$. Our experiments use no augmentation since the study focuses on sampling strategy performance. During the training phase, we use the AdamW optimizer with decay weights of $1 \times 10^{-5}$, a learning rate of $4 \times 10^{-4}$, and a batch size of $8$. After $40000$ iterations training, the model checkpoint with the highest Dice similarity score (DSC) in the validation set will be selected for inference. The dropout rate is $0.5$ with $T=10$ times MC dropouts during the training and inference phases.

\textbf{Acquisition rules.}
Selecting all sequential images from one volume (i.e., \textit{volume-wise} acquisition) may introduce redundant information to the training pool \cite{chen2023survey}, as neighboring images usually exhibit similar features. On the other hand, from an annotator's point of view, annotating an entire volume is more efficient than annotating an equal number of slices among different volumes (i.e., \textit{slice-wise} acquisition), as it requires less time to operate the software to locate the target slice, and the annotation of consecutive slices can leverage semi-automatic tools for the slice interpolation. In order to analyze this trade-off, all experiments are conducted under both slice-wise and volume-wise acquisition.

\textbf{Sampling strategies.}
We compared the proposed method with several recent AL algorithms in medical image segmentation tasks to demonstrate its efficacy and robustness in different scenarios. The chosen strategies for comparison include random selection, uncertainty-only selection \cite{gal2017deep}, and two state-of-the-art (SOTA) methods \cite{hiasa2019automated, nath2020diminishing}. The details and rationale for selecting these methods are as follows:

\noindent$\rightarrow$ $\text{BAL}_{rand}$: a simple baseline of randomly selecting samples from $\mathcal{D}_{u}$ illustrates the improvement of other methods over a non-strategic approach.

\noindent$\rightarrow$ $\text{BAL}_{unc}$: a common uncertainty-based AL approach that selects the most uncertain samples from $\mathcal{D}_{u}$ based on Section \ref{subsec4}.

\noindent$\rightarrow$ $\text{BAL}_{unc+sim}$: a combination of uncertainty and density-based representative sampling described in \cite{hiasa2019automated}.  $sim$ denotes the cosine similarity between the unlabeled pool and the selected samples, which is maximized for density enhancement. This approach represents a strong SOTA method, particularly in the segmentation of musculoskeletal structures.

\noindent$\rightarrow$ $\text{BAL}_{unc+mi}$: resembles the method described in \cite{nath2020diminishing}, which integrates uncertainty sampling with a diversity constraint, offering a robust comparison in the context of 3D medical image segmentation. In our implementation, we have adapted this approach by substituting the standard uncertainty calculation with Bayesian estimation and setting the ''Delete Flag'' to $1$.

\noindent$\rightarrow$ $\text{BAL}_{unc+hres}$: our proposed method that combines uncertainty sampling and hybrid representation-enhanced sampling (Section \ref{subsec4}), with the hyperparameter $\lambda$ empirically set to 0.5 and 0.25 for volume-wise and slice-wise acquisition, respectively. 

\begin{figure}[t]%
\centering
\includegraphics[width=1.0\textwidth]{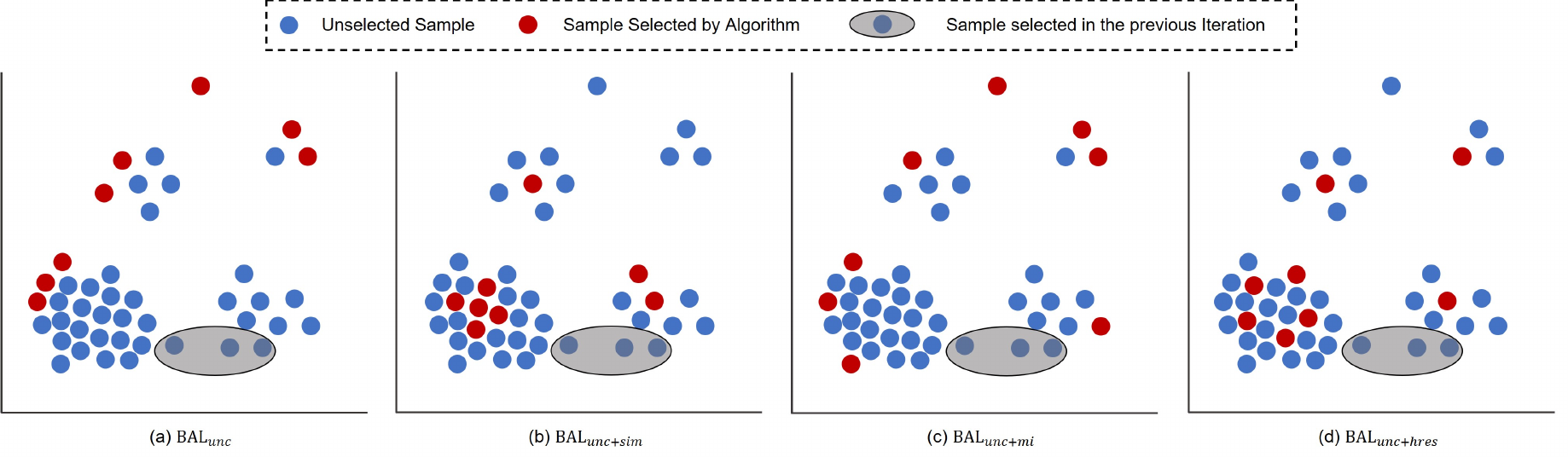}
\caption{Schematic visualization of sampling strategies on a theoretical 2D hyperplane. Unselected samples are marked in blue, while red points indicate those selected by the corresponding algorithm. The shaded regions delineate the samples previously chosen for model training. (a)  $\text{BAL}_{unc}$, (b) $\text{BAL}_{unc+sim}$, (c) $\text{BAL}_{unc+mi}$, (d) $\text{BAL}_{unc+hres}$ (the proposed method).}\label{fig5}
\end{figure}

To better demonstrate the operational mechanics of these strategies under idealized conditions, we present a comparative visualization of their selection processes in Fig. \ref{fig5}. $\text{BAL}_{unc}$ selects samples as distant as possible from the model's existing knowledge base. In contrast, $\text{BAL}_{unc+sim}$ chooses samples that both represent the overall sample density distribution and are distanced from the model's knowledge. $\text{BAL}_{unc+mi}$ aims to maximize the diversity among chosen samples, again focusing on those outside the current scope of the model's knowledge. Finally, $\text{BAL}_{unc+hres}$ aims to balance the representation of the density distribution and the internal diversity of samples, while selecting from areas not covered by the model's existing knowledge. Additionally, an ablation study was conducted to assess the individual contributions of these sampling strategies to the overall performance.

\textbf{Evaluation metrics}
The segmentation accuracy was assessed at each acquisition step using DSC. To quantify the manual labor saved by our BAL framework, we proposed a metric called $reduced$ $annotation$ $cost$ (RAC) as
\begin{equation}\label{eq3}
    RAC(I)=1-\frac{|I^{revised}|}{|I^{ROI}|}
\end{equation}
with the queried label image $I$. $| I^{revised}|$ denotes the number of pixels/voxels to be revised, whereas $| I^{ROI} |$ is the number of non-background pixels in the corresponding ground truth. Unlike the manual annotation cost (MAC) used in \cite{hiasa2019automated} that considers all image pixels, RAC considers non-background pixels, as annotation tools initially assign a zero value to all pixels and annotators modify only non-background ones. This methodology is consistent with common practices in multi-structure medical imaging, where experts frequently perform pixel-level or localized revisions. Such precision is particularly necessary for closely adjacent structures, where automated tools lack the needed granularity. In contrast to the percentage of labeled training data, which denotes the ratio of revised slices or volumes and is widely used in AL research \cite{yang2017suggestive, ozdemir2021active, li2023hal}, RAC measures the ratio of pixels requiring revision by annotators. Therefore, RAC more accurately captures the extensive manual revisions required in these scenarios, as it focuses on pixel/voxel-level analysis.

\section{Results}\label{sec4}
Comparative results on both MRI and CT datasets. t-distributed Stochastic Neighbour Embedding (t-SNE) maps and raw data of DSC and RAC for all strategies at each active iteration are available in Online Resource 1.4 and Online Resource 2. Additionally, we have made available tables showing structure-wise average model performance for two datasets and two acquisition methods in Online Resource 1.5.
\subsection{MRI dataset}\label{subsec7}
\begin{figure}[t]%
\centering
\includegraphics[width=1.0\textwidth]{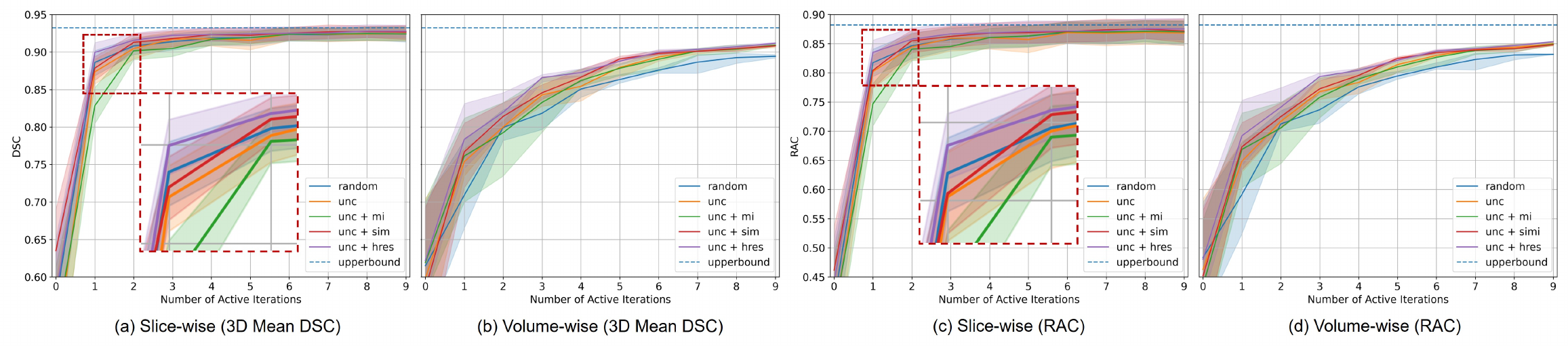}
\caption{DSC and RAC results on the MRI dataset. The upper bound (blue dashed line) denotes the average DSC/RAC when trained on all 90 volumes, while the purple line represents $\text{BAL}_{\text{unc+hres}}$ (proposed). (a) DSC for slice-wise acquisition, (b) DSC for volume-wise acquisition, (c) RAC for slice-wise acquisition, and (d) RAC for volume-wise acquisition. Each set of results includes a 95\% confidence interval, with (a) and (c) based on 20 testing samples and (b) and (d) on three random seeds.}\label{fig3}
\end{figure}

% \begin{table}[t]
% \begin{center}
% \resizebox{1.0\textwidth}{!}{%
% \begin{minipage}{1.05\textwidth}
% \caption{Comparison of reduced annotation cost (RAC) metric on MRI dataset presented as mean $\pm$ std \%, where a larger value indicates fewer pixels to be revised. Notice that the mean RAC of the upper bound is $88.2$\% as the model trained on the fully labeled dataset.}\label{tab2}
% \begin{tabular*}{\textwidth}{@{\extracolsep{\fill}}lcccccc@{\extracolsep{\fill}}}
% \toprule%
% & \multicolumn{6}{@{}c@{}}{Annotation\footnotemark[1]\% (number of active iteration)} \\\cmidrule{2-7}%
% RAC \% & 1.25 (0) & 2.50 (1) & 5.00 (3) & 7.50 (5) & 10.0 (7) & 12.5 (9)\\
% \midrule
% BAL$_{rand}$  & $\mathbf{48.3}\pm1.0$ & $59.1\pm0.6$ & $73.7\pm0.1$ & $79.5\pm0.0$ & $82.3\pm0.0$ & $83.2\pm0.0$\\
% BAL$_{unc}$  & $46.3\pm1.4$ & $64.9\pm0.1$ & $76.9\pm0.0$ & $81.4\pm0.0$ & $84.0\pm0.0$ & $85.0\pm0.0$\\
% BAL$_{unc+mi}$  & $43.5\pm1.5$ & $66.9\pm0.4$ & $75.8\pm0.1$ & $81.0\pm0.0$ & $83.8\pm0.0$ & $84.9\pm0.0$\\
% BAL$_{unc+sim}$  & $45.2\pm1.1$ & $67.3\pm0.2$ & $77.3\pm0.0$ & $\mathbf{82.5}\pm0.0$ & $83.9\pm0.0$ & $84.8\pm0.0$\\
% BAL$_{proposed}$  & $48.0\pm0.6$ & $\mathbf{69.2}\pm0.3$ & $\mathbf{79.4}\pm0.0$ & $82.1\pm0.0$ & $\mathbf{84.2}\pm0.0$ & $\mathbf{85.3}\pm0.0$\\
% \botrule
% \end{tabular*}
% \footnotetext{Exp. settings: volume-wise acquisition; 4-layer network of 2.73 million trainable parameters.}
% \footnotetext[1]{Percentile of the annotated data used for model training.}
% \end{minipage}
% }
% \end{center}
% \end{table}

Initialized with one randomly selected volume of 180 slices, we selected and revised one volume (for volume-wise selection) or 180 slices (average slice count per volume in the unlabeled pool, for slice-wise selection) from $\mathcal{D}_{u}$ and added them to $\mathcal{D}_{t}$ at each iteration. Data partitioning of volume-wise experiments was conducted three times with different random seeds to ensure reliability. The DSCs of all methods on the MRI dataset are shown in Fig. \ref{fig3}, where $\text{BAL}_{unc+hres}$ is depicted as the purple line. Comparing Fig. \ref{fig3} (a) with (b), we can infer that slice-wise acquisition systematically surpasses volume-wise by around 0.1 DSC, reaching close to the upper bound within five active iterations. Focusing on the method comparisons, the DSC of $\text{BAL}_{unc+hres}$ shows a modest improvement compared to the SOTA $\text{BAL}_{unc+sim}$ and $\text{BAL}_{unc+mi}$, in most iterations.

 In our evaluation of framework efficiency, we specifically assessed the RAC of different BAL methods, as shown in Fig. \ref{fig3} (c) and (d). These figures illustrate a trend consistent with the DSC results, underlining the efficiency of the various methods. Notably, our proposed method makes a moderate contribution to reducing the annotation cost, particularly in the early iterations. This efficiency is most pronounced in the 3rd active iteration (79.4\% in RAC), where our method surpasses the second-best approach, $\text{BAL}_{unc+sim}$, by an improvement of 2.1\% in RAC. This improvement underscores the effectiveness of our proposed method in enhancing both accuracy and annotation efficiency in quadriceps muscle segmentation from MRI images.

\subsection{CT dataset}\label{subsec8}
\begin{figure}[t]%
\centering
\includegraphics[width=1.0\textwidth]{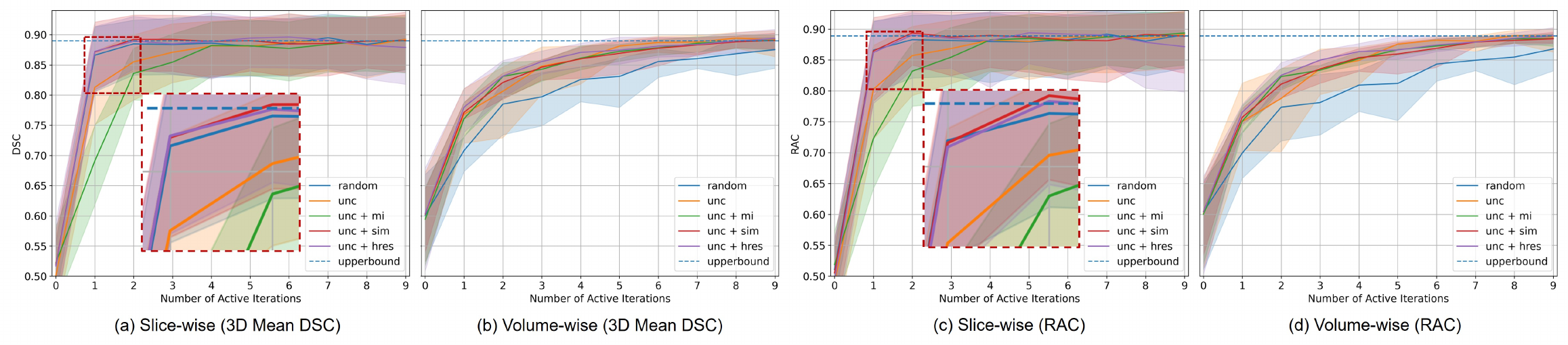}
\caption{DSC and RAC results on the CT dataset. The upper bound (blue dashed line) denotes the average DSC/RAC when trained on all 25 volumes, while the purple line represents $\text{BAL}_{\text{unc+hres}}$ (proposed). (a) DSC for slice-wise acquisition, (b) DSC for volume-wise acquisition, (c) RAC for slice-wise acquisition, and (d) RAC for volume-wise acquisition. Each set of results includes a 95\% confidence interval, with (a) and (c) based on 4 testing samples and (b) and (d) on three random seeds.}\label{fig4}
\end{figure}

Experiments on the CT dataset employed similar settings to those presented in Section. \ref{subsec7}, except that the slice-wise experiment chose 540 slices per iteration corresponding to the size of one volume in $\mathcal{D}_{u}$. The DSC results in Fig. \ref{fig4} show a similar trend to those obtained from the MRI dataset. A notable observation from Fig. \ref{fig4} (a) is the pronounced performance disparity between the group comprising $\text{BAL}_{rand}$, $\text{BAL}_{unc+hres}$, $\text{BAL}_{unc+sim}$, and the other methods. Surprisingly, random selection demonstrated unexpectedly strong performance in the slice-wise acquisition. When focusing on volume-wise acquisition (Fig. \ref{fig4} (b)), $\text{BAL}_{unc+hres}$ shows marginal gain to the rest combinations, especially at the early stages of the 1st and 3rd active iterations (both 0.01 improvement in DSC, over the second-ranking method). Fig. \ref{fig4} (c) and (d) illustrate the RAC results in line with those shown in Section \ref{subsec7}. Our method achieved the upper bound RAC using only 12\% and 40\% of the full training data with slice-wise and volume-wise acquisition, respectively. 

\subsection{Ablation study}\label{subsec9}
\begin{table}[t]
\begin{center}
\resizebox{\textwidth}{!}{%
\begin{minipage}{\textwidth}
\caption{Results of an ablation study on $\text{BAL}_{\text{unc+hres}}$ (proposed) components, showing mean (std) model performance across two datasets from the 1$^{\text{st}}$ to the 6$^{\text{th}}$ active iteration.}
\label{tab4}

\begin{tabular}{@{}l|c|c|c|cc|cc|cc|cc@{}}
\toprule
 & \multicolumn{3}{c|}{AL Selection} & \multicolumn{2}{c|}{MRI\footnotemark[1] (Vol)} & \multicolumn{2}{c|}{MRI\footnotemark[1] (Slice)} & \multicolumn{2}{c|}{CT\footnotemark[2] (Vol)} & \multicolumn{2}{c}{CT\footnotemark[2] (Slice)} \\ 
\cmidrule{2-12}
\multirow{-2}{*}{\textbf{Method}} & \textbf{UNC} & \textbf{MI} & \textbf{SIM} & \textbf{DSC} & \textbf{RAC} & \textbf{DSC} & \textbf{RAC} & \textbf{DSC} & \textbf{RAC} & \textbf{DSC} & \textbf{RAC} \\
\midrule
\multirow{2}{*}{$\text{BAL}_{\text{rand}}$} & \multirow{2}{*}{-} & \multirow{2}{*}{-} & \multirow{2}{*}{-} & 81.9$^{*}$ & 73.7 & 91.2$^{*}$ & 85.2$^{*}$ & 80.0$^{*}$ & 78.6$^{*}$ & 88.1 & 87.9 \\
 & & & & ($\pm$6.11) & ($\pm$7.99) & ($\pm$1.35) & ($\pm$1.87) & ($\pm$5.18) & ($\pm$4.93) & ($\pm$0.77) & ($\pm$0.64) \\
\midrule
\multirow{2}{*}{$\text{BAL}_{\text{unc}}$} & \multirow{2}{*}{\checkmark} & \multirow{2}{*}{-} & \multirow{2}{*}{-} & 83.7$^{*}$ & 76.0$^{*}$ & 90.9 & 84.9 & \textbf{84.2} & 83.0 & 86.3 & 86.5 \\
 & & & & ($\pm$5.22) & ($\pm$6.74) & ($\pm$1.84) & ($\pm$2.40) & ($\pm$4.61) & ($\pm$5.16) & ($\pm$2.73) & ($\pm$2.78) \\
\multirow{2}{*}{$\text{BAL}_{\text{unc+mi}}$} & \multirow{2}{*}{\checkmark} & \multirow{2}{*}{\checkmark} & \multirow{2}{*}{-} & 83.6$^{*}$ & 76.0$^{*}$ & 89.9 & 83.8 & 84.1 & \textbf{83.3} & 83.7 & 84.3 \\
 & & & & ($\pm$5.08) & ($\pm$6.19) & ($\pm$3.57) & ($\pm$4.59) & ($\pm$4.32) & ($\pm$4.44) & ($\pm$7.40) & ($\pm$6.23) \\
\multirow{2}{*}{$\text{BAL}_{\text{unc+sim}}$} & \multirow{2}{*}{\checkmark} & \multirow{2}{*}{-} & \multirow{2}{*}{\checkmark} & \textbf{84.7} & \textbf{77.1} & \textbf{91.3} & \textbf{85.5} & 84.1$^{*}$ & 83.1$^{*}$ & \textbf{88.7} & \underline{\textbf{88.4}} \\
 & & & & ($\pm$4.99) & ($\pm$6.21) & ($\pm$1.76) & ($\pm$2.55) & ($\pm$3.94) & ($\pm$4.16) & ($\pm$0.79) & ($\pm$1.02) \\
\midrule
\multirow{2}{*}{$\text{BAL}_{\text{unc+hres}}$\footnotemark[3]} & \multirow{2}{*}{\checkmark} & \multirow{2}{*}{\checkmark} & \multirow{2}{*}{\checkmark} & \underline{\textbf{85.5}} & \underline{\textbf{78.2}} & \underline{\textbf{91.8}} & \underline{\textbf{86.2}} & \underline{\textbf{84.9}} & \underline{\textbf{84.1}} & \underline{\textbf{88.8}} & \underline{\textbf{88.4}} \\
 & & & & ($\pm$4.44) & ($\pm$5.47) & ($\pm$0.98) & ($\pm$1.40) & ($\pm$3.81) & ($\pm$4.07) & ($\pm$0.85) & ($\pm$1.18) \\
\botrule
\end{tabular}

\footnotetext[1]{MRI data: The quadriceps dataset annotated in active iterations ranging from 2.5\% (the 1$^{\text{st}}$) to 8.3\% (the 6$^{\text{th}}$).}
\footnotetext[2]{CT data: The musculoskeletal dataset annotated in active iterations ranging from 8.0\% (the 1$^{\text{st}}$) to 28.0\% (the 6$^{\text{th}}$).}
\footnotetext[3]{The proposed method integrates hybrid representativeness selection.}
\footnotetext {Note: \textbf{UNC}, \textbf{MI}, and \textbf{SIM} refer to the uncertainty module, diversity-enhanced module by mutual information, and density-enhanced module by cosine similarity, respectively. In this table, $^{*}$ denotes statistical significance after Bonferroni correction. Additionally, a bold value represents the second-highest value in the respective column for each metric, while a value in bold and underlined signifies the highest value.}
\end{minipage}
}
\end{center}
\end{table}

To illustrate the contribution of each component to $\text{BAL}_{unc+hres}$, we performed an ablation study that assesses the average model performance across the 1st to 6th active iterations, shown in Table \ref{tab4}. This range was chosen because, beyond the 6th iteration, there is a convergence in performance among different methods, making the inclusion of later iterations less meaningful for comparison. Our statistical analysis utilized the Bonferroni correction to adjust p-values for each set of method comparisons. These sets are defined by specific combinations: dataset type (either MRI or CT) and acquisition approach (volume-wise or slice-wise), evaluated under the DSC or RAC metrics.

The results generally indicate that integrating uncertainty with hybrid representativeness sampling yields modest improvements over other combinations. Notably, the paired t-test results indicate significant statistical differences in volume-wise acquisition, whereas the significance is less pronounced in slice-wise acquisition. Upon the comparison of $\text{BAL}_{rand}$ and $\text{BAL}_{unc}$, the uncertainty-based sampling significantly contributes 1.8\% and 4.2\% of DSC, and 2.3\% and 4.4\% of RAC, in the volume-wise acquisitions of MRI and CT datasets, respectively. Nevertheless, this trend reverses in slice-wise acquisition. The results for $\text{BAL}_{mi}$ and $\text{BAL}_{sim}$ indicate the impact of enhanced diversity and density. Compared to $\text{BAL}_{unc}$, solely incorporating an MI-based diversity regularizer can deteriorate the BAL performance. However, $\text{BAL}_{unc+hres}$ suggests that integrating the regularizer with a density-enhanced module effectively counteracts its negative impact, as this combination tends to select fewer redundant samples or outliers.

\section{Discussion}\label{sec5}
 % Since our sampling algorithm (Online Resource 1. b) demonstrates sensitivity to increment size, the computational expense of slice-wise acquisition is exponentially higher.
We proposed a hybrid representation-enhanced sampling strategy in BAL by integrating density-based and diversity-based criteria and evaluated its performance on MRI and CT datasets. Remarkably, our method achieves comparable performance to models trained on the full dataset using only a fraction of the data\textemdash10\% in the MRI dataset and 24\% in the CT dataset. This efficiency represents a modest yet robust improvement over other BAL methods using an equivalent number of training samples. Furthermore, we proposed a new metric, RAC (Ratio of Annotation Correction), for the quantitative estimation of annotation effort.

One can infer from Fig. \ref{fig3} and \ref{fig4} that $\text{BAL}_{unc+hres}$ show modest improvement over two SOTA samplings in the early iterations, though this improvement consistently decreases over iterations. One possible explanation is that the proposed method $\text{BAL}_{unc+hres}$ identified the key samples on both datasets ahead of other methods. Additionally, we observed that as the selection granularity shifts from volume-wise to slice-wise, the performance gap between the proposed method and other techniques narrows, as demonstrated by the statistical analysis presented in Table \ref{tab4}. Particularly, $\text{BAL}_{unc}$ and $\text{BAL}_{unc+mi}$ underperform in the initial iterations (Fig. \ref{fig4} (a) and (c)), likely due to the skewness in data selection with higher granularity and the limited data range of the CT dataset, respectively. The great performance of $\text{BAL}_{rand}$ in slice-wise acquisition is also in line with the findings in \cite{nath2020diminishing, GAILLOCHET2023102958}. Compared to the SOTA methods $\text{BAL}_{unc+sim}$ and $\text{BAL}_{unc+mi}$, our hybrid representation-enhanced sampling mitigates the overlap of information between samples and outliers in a robust way, as shown by Table \ref{tab4}. These results highlight the advantages of integrating a density and diversity-based scheme in BAL, improving segmentation accuracy in low-label datasets and reducing the required training samples. Our findings demonstrate that incorporating density, diversity, and uncertainty enhances segmentation accuracy, aligning with results in non-medical domains \cite{yuan2019multi} and are consistent with findings in medical classification \cite{liu2022survey}, indicating its broad applicability.

In both MRI and CT datasets, we observe similar trends in the RAC results as in the DSC ones. Table \ref{tab4} reveals that although the MRI and CT datasets show comparable DSC in volume-wise acquisition, the RAC difference between them can be as high as 7.3\%. The variation might be because mean DSC was used for multi-class segmentation, while RAC focused solely on misclassified pixels unaffected by the number of classes (4 structures in the case of MRI and 22 for CT scans). This approach provides valuable insights into the estimation of annotation costs. Moreover, the comparisons between (c) and (d) in both Fig. \ref{fig3} and \ref{fig4} illustrate that even with an equal percentage of annotated training data (number of active iterations), the annotation effort required varies significantly between datasets. This variation in annotation effort underscores the need for tailored strategies in musculoskeletal structure annotation, ensuring efficiency and accuracy across different medical imaging modalities. The impact of volume-wise and slice-wise acquisition on model performance and annotation cost is further discussed in Online Resource 1.6.

Despite showing some moderate improvement over alternative approaches, our study shows several limitations. Firstly, we implemented our hybrid scoring approach using a greedy algorithm with a computational cost of $O(n^3)$. This cost escalates exponentially with larger dataset sizes and finer acquisition granularity, limiting the method's effectiveness in large-scale datasets comprising thousands of image volumes. Secondly, we have limited our analysis of the impact of two acquisition rules on LE datasets, while alternative conclusions may be reached on other datasets. Finally, our AL sampling strategy, which excelled by leveraging the most valuable samples with only roughly 5-16\% of the training data, could further boost accuracy by tapping into the potential of the remaining unlabeled data without needing extra annotations. Thus, future works shall include 1) algorithm optimization (e.g., VAE-based measures) for efficiency improvement, 2) extensive experiments on various datasets for quantitative estimation of acquisition rules' impact, and 3) incorporating the semi-supervised learning (SSL) to unleash the potential of unlabeled data. Implementing SSL during the segmentation stage of each active iteration could significantly boost the model’s segmentation accuracy by utilizing unlabeled data as low-confidence training data \cite{nath2022warm}.

\section{Conclusion}\label{sec6}
This paper has described a BAL framework based on Bayesian U-net that leverages the advantage of AL to reduce annotation efforts. At the algorithmic level, we introduced a novel hybrid representation-enhanced sampling that ensures high density and diversity of the training data to boost the BAL framework's performance. Moreover, we conducted a comprehensive study to reveal the impact of acquisition rules on BAL, as well as parameter sweeping for a real-world clinical setting. The experiment results indicated that our proposed sampling strategy shows a moderate improvement over SOTA representativeness-based sampling approaches on musculoskeletal segmentation. However, comparison experiments between the two acquisition strategies indicate that the improvement diminishes as the granularity of the selected samples increases. We also summarized previous works for a better comprehension of our experiments (Online Resource 1.3), and our code is available on GitHub.\footnotemark[1]
\backmatter

\bmhead{Supplementary information}
Online Resource 1:  1) Details of Estimation of model uncertainty, 2) the hybrid representation-enhanced sampling algorithm, 3) a table summarizing previous works, 4) t-SNE maps, 5) tables presenting the average model performance by musculoskeletal structure for MRI and CT datasets, along with volume-wise and slice-wise acquisition strategies, and 6) a discussion of the two acquisition strategies and their relation to annotation costs.
Online Resource 2: Raw data of DSC and RAC for all methods and active iterations, available on GitHub.\footnotemark[1]
\footnotetext[1]{\url{https://github.com/RIO98/Hybrid-Representation-Enhanced-Bayesian-Active-Learning}.}
\bmhead{Acknowledgments}

This work was funded by MEXT/JSPS KAKENHI (19H01176, 20H04550, 20K19376, 21H03303, 21K16655, 21K18080).

% \section*{Declarations}
% \noindent\textbf{Ethics approval} Ethical approval was obtained from the Institutional Review Boards (IRBs) of the institutions participating in this study (IRB approval numbers: 21115 for Osaka University Hospital, R1746-2 for Kyoto University Hospital, and 2020-M-7 for Nara Institute of Science and Technology.)


\bibliography{sn-bibliography}% common bib file

\end{document}